\documentclass{tMPH2e}
\usepackage{color}
\usepackage{graphicx}
\usepackage[usenames,dvipsnames]{xcolor}
\usepackage{hyperref}

\newcommand{\cm}{cm$^{-1}$}
\newcommand{\leftexp}[2]{{\vphantom{#2}}^{#1}{#2}}   

\usepackage{verbatim}

\begin{document}
\doi{10.1080/0026897YYxxxxxxxx}
\jvol{00}
\jnum{00} \jyear{2015} 
\title{The calculated rovibronic spectrum of scandium hydride, ScH}

\author{Lorenzo Lodi, Sergei N. Yurchenko and Jonathan Tennyson\\
\vspace{6pt} {\em{Department of Physics \& Astronomy, University College London,  London WC1E~6BT, United Kingdom}} \\
\vspace{6pt}\received{received \today} }

\maketitle

\begin{abstract}
  The electronic structure of six low-lying electronic states of scandium
  hydride, $X\,{}^{1}\Sigma^+$, $a\,{}^{3}\Delta$, $b\,{}^{3}\Pi$,
  $A\,{}^{1}\Delta$ $c\,{}^{3}\Sigma^+$, and $B\,{}^{1}\Pi$, is studied using
  multi-reference configuration interaction as a function of bond length.
  Diagonal and off-diagonal dipole moment, spin-orbit coupling and electronic
  angular momentum curves are also computed. The results are benchmarked
  against experimental measurements and calculations on atomic scandium.  The
  resulting curves are used to compute a line list of molecular ro-vibronic
  transitions for $^{45}$ScH.

\noindent {\bf{Acknowledgment}}
This work was supported by the ERC under Advanced Investigator Project 267219.
\end{abstract}

\section{Introduction}\label{sec:intro}

Scandium hydride was first identified experimentally by Smith
\cite{73Smxxxx.ScH}, who recorded the absorption spectra of
various transition metal hydrides (ScH, TiH, VH, NiH, CoH
and deuterated isotopologues) in the region 17700 to 18300~\cm.
No detailed analysis of the spectrum was reported but it was remarked
that a triplet ground state was expected.
Later studies showed that the
ground state of ScH is actually $\leftexp{1}{\Sigma^+}$,
with a low-lying  $\leftexp{3}{\Delta}$. 
Early studies using restricted open-shell Hartree-Fock (ROHF) \cite{74ScRixx.ScH,76ScRixx.ScH},
generalised valence bond theory \cite{75KuGuBl.ScH} and empirically-fitted pseudopotentials \cite{81Daxxxx.ScH}
all incorrectly predicted a $\leftexp{3}{\Delta}$ ground state, with the $\leftexp{1}{\Sigma^+}$
generally lying about 2000~\cm{} higher up. These studies considered the six electronic terms
correlating upon dissociation with ground state atoms (dissociation channel
labelled 1 in table~\ref{tbl.Sc.atom.experiment}, leading to $\leftexp{1,3}{\Sigma}$, $\leftexp{1,3}{\Pi}$ and $\leftexp{1,3}{\Delta}$);
it was remarked  \cite{75KuGuBl.ScH} that the bonding of the molecular terms
other than $\leftexp{1}{\Sigma^+}$ is due
to the Sc($4s$) and H($1s$) orbitals, while the scandium $3d$ orbitals are
relatively unaffected with respect to the atomic state \cite{11Hougen}.
On the other hand the bonding of $\leftexp{1}{\Sigma^+}$ was
put down to $spd$ bonding \cite{75KuGuBl.ScH,79Pyxxxx.ScH}.
The different bonding character of the $\leftexp{1}{\Sigma^+}$ term is probably one of
the reasons
why an extensive treatment of electron correlation is necessary to obtain the right ordering
of the electronic terms.
Bauschlicher and  Walch \cite{82BaWaxx.ScH} were the first to correctly predict
$\leftexp{1}{\Sigma^+}$ lying below $\leftexp{3}{\Delta}$ 
by performing full-valence multi-configuration self-consistent-field (MCSCF) calculations.
Jeung and Kouteck{\'y} \cite{88JeKoxx.ScH} studied the same six electronic terms
using pseudopotentials and truncated MRCI and confirmed a ground $\leftexp{1}{\Sigma^+}$ term close
to equilibrium ($r_e \approx 3.4~\mathrm{a}_0$), although for longer bond lengths
$\leftexp{3}{\Delta}$ becomes lower in energy.
Note that all these studies kept the scandium outer-core $3s3p$ electrons uncorrelated and often
did not include relativistic corrections. 

Anglada \emph{et al} produced a series of papers \cite{83AnBrPe.ScH,84AnBrPe.ScH,86BrAnxx.ScH,89AnBrPe.ScH}
studying in great detail ScH and ScH$^+$ using multi-reference configuration interaction (MRCI);
in particular their final paper \cite{89AnBrPe.ScH} constitutes the most
complete theoretical study of ScH currently available. These authors confirmed that $\leftexp{1}{\Sigma^+}$
is the ground state when correlation effects are included; they also found that
correlation of the Sc($3s 3p$) semi-core electrons leads to large energy shifts, strongly
stabilising the $\leftexp{1}{\Sigma^+}$ term with respect to the others
and swapping the order of some of the excited states. They used basis sets
similar in size to cc-pVTZ.

More recent theoretical studies considered ScH in the context of calibration
studies of transition metal molecules using density function theory
\cite{01GuGoxx.ScH,08GoMaxx.ScH}, but they focused on equilibrium properties of
very few terms and are of little relevance for us. An exception is the very recent
study by Hubert \emph{et al} \cite{13HuOlLo.ScH} in which a detailed study of
the ground $\leftexp{1}{\Sigma^+}$ and of two excited terms
$\leftexp{1,3}{\Delta}$ around equilibrium was presented and a modification of
the coupled cluster method called general active space coupled cluster (GAS-CC)
was used.

The only theoretical dipole moment data available for ScH are those by
Anglada \emph{et al}  \cite{89AnBrPe.ScH} and by Chong \emph{et al} \cite{86ChLaBa.ScH}.

Experimentally a study by Bernard \emph{et al} \cite{77BeEfBa.ScH} reported
three new bands ascribed to ScH and ScD in the region 11600 to 12700~\cm, but
no detailed analysis or assignments were made due to the limited resolution
and complexity of the spectra.
More recently Ram and Bernath \cite{96RaBexx.ScH,97RaBexx.ScH} reported two
detailed emission spectra analyses for ScH and ScD. In these studies they
reported on singlet-singlet bands
in regions from 5400 to 20500~\cm{} assigned to transitions between 8
electronic terms, namely $X\,{}\leftexp{1}{\Sigma^+}$,
$A\,{}\leftexp{1}{\Delta}$, $B\,{}\leftexp{1}{\Pi}$,
$C\,{}\leftexp{1}{\Sigma^+}$, $D\,{}\leftexp{1}{\Pi}$,
$E\,{}\leftexp{1}{\Delta}$, $F\,{}\leftexp{1}{\Sigma^-}$ and
$G\,\leftexp{1}{\Pi}$. Two additional strong bands near 11620 and 12290~\cm{}
and two weaker bands near 12660 and 16845~\cm{} were recorded but only
incompletely analysed; the band near 11620~\cm{} was conjectured to be due to a
transition to the low-lying $\leftexp{3}{\Delta}$ term from a
$\leftexp{3}{\Phi}$ term.
Le and Steimle \cite{11LeStxx.ScH} reported more recently a detailed
experimental study of the $X\,{}\leftexp{1}{\Sigma^+}$--$D\,{}\leftexp{1}{\Pi}$
band around 16850~\cm, where also the electric dipole moments of ScH in its
$X\,{}\leftexp{1}{\Sigma^+}$ and $D\,\leftexp{1}{\Pi}$ states were obtained
using optical Stark spectroscopy. Very recently,  Mukund {\it et al}
\cite{14MoBhNa.ScH} reported the observation of ScH emission bands at about
17900~\cm, ascribed to the $g$~$\leftexp{3}{\Phi}$ -- $a$~$\leftexp{3}{\Delta}$
triplet-triplet transitions.

This paper focuses on the six low-energy electronic states dissociating to
ground state Sc and H atoms. Of the various experimentally observed
bands \cite{96RaBexx.ScH,97RaBexx.ScH,11LeStxx.ScH} only $X$ -- $B$ is
considered in the present study, although experiment is used for the empirical
refinement of the potential energy curves of the singlet terms
$X\,{}\leftexp{1}{\Sigma}$, $A\,{}\leftexp{1}{\Delta}$ and
$B\,{}\leftexp{1}{\Pi}$ as well.  

As part of the ExoMol project \cite{jt528}, whose aim is to produce comprehensive line lists
for hot, astrophysically-important molecules, we have been constructing rovibrational and rovibronic
line lists for a number of diatomic species \cite{jt529,jt563,jt583,jt590,jt598}. However, so
far none of these have contained a transition metal (TM). The richness of the spectrum of TM-containing
diatomics makes their opacity particularly important for astrophysical studies \cite{aha97} but
treating their rovibronic spectrum {\it ab initio} is very challenging.
Scandium hydride is the lightest TM
molecule and for this reason constitutes a useful
benchmark system for theoretical studies of such systems.

Scandium hydride has received comparatively little attention with respect to
other transition metal hydrides such as FeH or NiH, in all probability because
of the low abundance of scandium. Scandium is in fact the rarest of
fourth-period transition metals (Sc-Zn) in the solar system \cite{03Lodders},
although it is more abundant than all heavier elements starting from the fifth
period. The study presented here on ScH constitutes a first step in the {\it ab
initio} calculation of ro-vibronic spectra of TM-containing molecule. We
perform a series of {\it ab initio} calculations on both the scandium atom and
ScH and use these to produce a line list of scandium hydride line positions and
intensities which is reasonably complete in the region up to 12~000~\cm.

\section{Atomic Scandium}
As a preliminary test we studied in some detail the scandium atom
using complete active space self consistent field (CASSCF) and internally-contracted
MRCI calculations using the program Molpro \cite{12WeKnKn.methods}.

We collected in table~\ref{tbl.Sc.atom.experiment} reference energy levels of
the scandium atom up to about 20~000~\cm, along with the ScH molecular terms
correlating adiabatically with the various atomic states; this information
serves as an indication of which molecular terms are expected to be low-lying;
the rationale is that for transition-metal hydrides the hydrogen atom
constitutes a relatively small perturbation of the atomic energy levels, so
that molecular terms correlating with high-energy atomic products should be
high-lying too. The lowest dissociation channel is separated by about
11~500~\cm{} from others, and in fact the six electronic terms correlating to
it are lowest-lying and most theoretical studies concentrated on them. The
dissociation channels 1 to 10 reported in table~\ref{tbl.Sc.atom.experiment}
lead altogether to 60 electronic molecular terms; that is, there are 60 energy
curves in absence of
spin-orbit splitting, which become 155 
in presence of spin-orbit splitting. 
Of these, only eight have been characterised experimentally \cite{97RaBexx.ScH}.
Very probably many of the remaining molecular energy curves are repulsive (i.e., have no minimum)
and therefore do not concern us here.
In any case great complexity and strong perturbations in the observed spectra are expected.
This is typical of open-shell transition metal molecules.

In the following we consider the excitation energies from the ground
$\leftexp{2}{\mathrm{D}}$ term to the two lowest-energy excited terms, namely
$\leftexp{4}{\mathrm{F}}$ and $\leftexp{2}{\mathrm{F}}$.
We expect that errors of computed molecular excitation energies
are comparable with the corresponding error in the atomic case.

\begin{table*}
\begin{center}
\caption{Energy levels for scandium atom up to 20~200~\cm. The $\langle E \rangle$ are term-averaged energies, computed
by $\langle E \rangle = \frac{\sum_J (2J+1) E_J}{\sum_J (2J+1)}$;  experimental energies for the levels $E_J$ where taken
from the NIST website \cite{NISTWebsite}. $n_f=\min(2S+1,2L+1)$ is the number of fine-structure components of a given
terms due to spin-orbit interaction; $n_\mu=(2S+1)(2L+1)$ is the total degeneracy of the term (number of microstates).
The $\xi$'s are effective spin-orbit coupling constants, such that the spin-orbit splittings for each term are best
reproduced by the expression $ E_J = E_0 + \xi [(J(J+1)-L(L+1)-S(S+1)]/2 $, $J=|L-S|,\cdots,L+S$.
The last column lists the molecular terms for ScH correlating at dissociation with the given Sc atomic term plus a
ground state $^2\mathrm{S}$ hydrogen atom. \label{tbl.Sc.atom.experiment} }
\begin{tabular}{r l l rr r r l}
\hline
\hline
\# & config.   &  \multicolumn{1}{c}{term}         &  \multicolumn{1}{c}{ $\langle E \rangle$ / \cm}&$n_f$  &$n_\mu$& \multicolumn{1}{c}{ $\xi$ / \cm }    & \multicolumn{1}{c}{ molecular terms}  \\
\hline
1&$3d^1 4s^2$ & $\leftexp{2}{\mathrm{D}}$          &        0.0             &    2        &  10      &      67.3                &  $^{1,3}[\Sigma^+, \Pi, \Delta]$         \\
2&$3d^2 4s^1$ & $\leftexp{4}{\mathrm{F}}$          &   11~509.1             &    4        &  28      &      15.0                &  $^{3,5}[\Sigma^-, \Pi, \Delta,\Phi]$    \\
3&$3d^2 4s^1$ & $\leftexp{2}{\mathrm{F}}$          &   14~891.3             &    2        &  14      &      33.1                &  $^{1,3}[\Sigma^-, \Pi, \Delta,\Phi]$    \\
4&$3d^1 4s^1 4p^1 $ &  $\leftexp{4}{\mathrm{F^\circ}}$ &   15~775.8             &    4        &  28      &      33.6                &  $^{3,5}[\Sigma^+, \Pi, \Delta,\Phi]$            \\
5&$3d^1 4s^1 4p^1 $ &  $\leftexp{4}{\mathrm{D^\circ}}$ &   16~031.0             &    4        &  20      &      25.2                &  $^{3,5}[\Sigma^-, \Pi, \Delta]$            \\
6&$3d^1 4s^1 4p^1 $ &  $\leftexp{2}{\mathrm{D^\circ}}$ &   15~951.4             &    2        &  10      &     -29.7                &  $^{1,3}[\Sigma^-, \Pi, \Delta]$            \\
7&$3d^2 4s^1  $     &  $\leftexp{2}{\mathrm{D}}  $ &   16~916.7             &    2        &  10      &      -5.0                &  $^{1,3}[\Sigma^+, \Pi, \Delta]$            \\
8&$3d^2 4s^1  $     &  $\leftexp{4}{\mathrm{P}}  $ &   17~175.2             &    3        &  12      &      20.1                &  $^{3,5}[\Sigma^-, \Pi]$             \\
9&$3d^1 4s^1 4p^1 $ &  $\leftexp{4}{\mathrm{P^\circ}}$ &   18~440.6             &    3        &  12      &      15.0                &  $^{3,5}[\Sigma^+, \Pi]$             \\
10&$4s^2 4p^1 $ &  $\leftexp{2}{\mathrm{P^\circ}}$     &   18~706.5             &    2        &   6      &      96.5                &  $^{1,3}[\Sigma^+, \Pi]$             \\
\hline
\hline
\end{tabular}
\end{center}
\end{table*}

The most accurate study of (neutral or singly-ionized) transition metal atoms
including scandium is due to Balabanov and Peterson
\cite{05BaPexx.ScH,06BaPexx.ScH}. These authors used coupled cluster up to
CCSDTQ, relativistic corrections based on the Douglas-Kroll-Hess (DKH)
hamiltonian, included core correlation and developed the largest basis sets
available for transition metals. For scandium $\leftexp{2}{\mathrm{D}} \to
\leftexp{4}{\mathrm{F}}$ excitation energy the best theoretical coupled cluster
result is higher than the experimental one by about 115~\cm. The corresponding
result using ACPF (a multi reference method very close to MRCI) in conjunction
with the full-valence reference space is too low by about 110~\cm; using a
larger reference space including a further set of diffuse $d$ functions (which
are thought to be necessary for describing the late transition metals Fe-Cu)
leads to a worse agreement with experiment of about 190~\cm. Other recent
studies of transition metal atoms excitation energies including scandium were
performed by Raab and Roos \cite{05RaPoxx.ScH} and Mayhall \emph{et al}
\cite{08MaRaRe.ScH}.
Raab and Roos \cite{05RaPoxx.ScH}
computed the $\leftexp{2}{\mathrm{D}} \to \leftexp{4}{\mathrm{F}}$ excitation energy
with CCSD(T) and CASPT2 using the DKH hamiltonian for relativistic effects and the ANO-RCC basis set
(similar in size to aug-cc-pCVQZ). Both CCSD(T) and CASPT2 frozen core values agree
with experiment to about 250~\cm, but allowing for core correlation worsens somewhat the agreement to about
500~\cm. Mayhall \emph{et al} \cite{08MaRaRe.ScH} also used core-correlated CCSD(T) with the G3Large basis set
(similar in size to aug-cc-pCVTZ) and reported an agreement of 250~\cm{}
without inclusion of relativistic effects and of about 1200~\cm{} when
relativistic effected were included.

Table~\ref{tbl.Sc.atom.experiment} gives some indicative result for both the
$\leftexp{2}{\mathrm{D}} \to \leftexp{4}{\mathrm{F}}$ and
$\leftexp{2}{\mathrm{D}} \to \leftexp{2}{\mathrm{F}}$ transitions performed in
this study using MRCI and the full valence reference space.
Our best results for both transitions are too small on the average by about 750~\cm{} 
with respect to the results by Balabanov and Peterson; our errors are larger probably because
we performed a state-averaged calculation at the CASSCF level and also
because of the smaller basis sets used. A striking consideration is that the non-relativistic CASSCF
excitation energies are extremely good, a fact which can only be due to fortuitous cancellation
of errors. Overall our results and the analysis of the literature show that it is difficult
to get excitation energies correct to better than about 500~\cm, and that good agreement with experiment
can be often due to cancellation effects. We also observed that the relativistic contribution to excitation
energies shows relatively large variations of the order of 500~\cm{} depending on the
levels of electron correlation (CASSCF, MRCI valence only or core-correlated) and on
using the mass-velocity one-electron Darwin (MVD1) rather than the Douglas-Kroll-Hess (DKH) Hamiltonian.

\begin{table*}
\begin{center}
 \caption{Electronic term excitation energies for scandium atom (this work). All calculations used the full-valence (3-electron, 9-orbital) complete active space comprising the $3d4s4p$ orbitals. Orbitals are state-averaged over the three electronic terms considered. MRCI+Q are Davidson-corrected energies (relaxed reference). Calculated values are reported as (reference -- calculated),
 and reference energies are taken from the column labelled `$\langle E \rangle$' of table \ref{tbl.Sc.atom.experiment}. All quantities are in \cm. \label{tbl.Sc.atom.3z}}
\begin{tabular}{l l r r }
\hline
\hline
Method                 &    &$\leftexp{2}{\mathrm{D}} \to \leftexp{4}{\mathrm{F}}$ & $\leftexp{2}{\mathrm{D}} \to \leftexp{2}{\mathrm{F}}$ \\
\multicolumn{2}{r}{reference energies=}      &   11~509.1 & 14~891.3 \\
\mbox{}\\
      &     & \multicolumn{2}{c}{ref - calc}\\
\hline
CASSCF         & 3z  &      28.6  &   64.0  \\
CASSCF         & 4z  &      49.2  &   86.0  \\
\mbox{}\\
MRCI/frz core  & 3z   &   -1187.9  &  -421.8  \\
MRCI/frz core  & 4z   &   -1084.6  &  -265.3   \\
\mbox{}\\
MRCI/core corr  & wc3z   &  573.3  &   541.1  \\
MRCI/core corr  & wc4z   & 1072.1  &  1056.0  \\
MRCI+Q/core corr& wc4z   & 2080.0  & 2039.8   \\
MRCI+Q/core corr/DKH4 & wc4z-DK   & 813.6  &  707.1  \\
\hline
\hline
\end{tabular}
\end{center}
\end{table*}

We also computed atomic spin-orbit splitting constants using CASSCF and MRCI wave functions
as implemented in MOLPRO.
Results are collected in table~\ref{tbl.Sc.atom.SO}.
Spin orbit splitting constants show weak sensitivity to the size of the basis set
and already with the smallest 2z basis set are converged within 1~cm$^{-1}$.
The dependence on the electron correlation treatment is also weak, with the ground $\leftexp{2}{\mathrm{D}}$ term being
the most sensitive. Going from CASSCF to frozen-core MRCI increases $\xi(\leftexp{2}{\mathrm{D}})$ by +18~cm$^{-1}$ but reduces
$\xi(\leftexp{4}{\mathrm{F}})$ and $\xi(\leftexp{2}{\mathrm{F}})$ by only 1.5 and 0.9~cm$^{-1}$ respectively.
Correlating the $(3s3p)$ outer core increases $\xi(\leftexp{2}{\mathrm{D}})$ by 5~cm$^{-1}$, $\xi(\leftexp{4}{\mathrm{F}})$ by 1.8~cm$^{-1}$ and $\xi(\leftexp{2}{\mathrm{F}})$ by 2.6~cm$^{-1}$.
With respect to the experimentally-derived values we do not observe a clear pattern of
convergence with respect to the level of theory used, and the simplest
CASSCF/2z values agree with experiment practically as well as the core-correlated, large basis set MRCI ones.

Considering that errors of $\approx 5$~cm$^{-1}$ in spin-orbit couplings
are very small with respect to the error in the main non-relativistic energies
we conclude that it is quite acceptable to compute spin-orbit couplings
at a low level of theory.

\begin{table*}
\begin{center}
 \caption{Calculated spin-orbit constants $\xi$ for scandium atom. The column labelled `obs.' are
 experimentally derived values from table~\ref{tbl.Sc.atom.experiment}.
 All values are in \cm. \label{tbl.Sc.atom.SO}}
\begin{tabular}{r r r r r r r}
\hline
\hline
                               &      & \multicolumn{5}{c}{CASSCF}                          \\
transition                     & Obs. &   2z &   3z     &  4z      &    5z   \\ 
$\xi(\leftexp{2}{\mathrm{D}})$ & 67.3 &   57.5    &  57.6   &   57.9    &  57.9  \\ 
$\xi(\leftexp{4}{\mathrm{F}})$ & 15.0 &   18.5    &  18.5   &   18.6    &  18.6  \\ 
$\xi(\leftexp{2}{\mathrm{F}})$ & 33.1 &   37.2    &  37.2   &   37.4    &  37.4  \\ 
\mbox{}\\
                               &       & \multicolumn{5}{c}{MRCI (frz. core)}                 \\
transition                     &       &   2z       &   3z     &  4z   &    5z   \\ 
$\xi(\leftexp{2}{\mathrm{D}})$ & 67.3  &    74.7   &    75.6  &   76.0  &   76.0 \\ 
$\xi(\leftexp{4}{\mathrm{F}})$ & 15.0  &    16.8   &    17.0  &   17.1  &   17.1 \\ 
$\xi(\leftexp{2}{\mathrm{F}})$ & 33.1  &    35.9   &    36.4  &   36.5  &   36.6 \\ 
\mbox{}\\
                               &       & \multicolumn{5}{c}{MRCI (core correlated)}       \\
transition                     &       &           &   wc3z     &  wc4z &    wc5z \\ 
$\xi(\leftexp{2}{\mathrm{D}})$ & 67.3  &           &   80.3     &  81.2 &    81.3 \\ 
$\xi(\leftexp{4}{\mathrm{F}})$ & 15.0  &           &   18.7     &  18.9 &    19.0 \\ 
$\xi(\leftexp{2}{\mathrm{F}})$ & 33.1  &           &   38.6     &  39.1 &    39.2 \\ 
\hline
\hline
\end{tabular}
\end{center}
\end{table*}

\section{ScH molecule}
As discussed in the introduction and hinted by the results for the Sc atom
presented in the previous section, from the point of view of high-resolution
spectroscopy the accuracy expected for transition metal diatomics is much lower
than the one generally achievable for molecules made up by main-group atoms.
Also, because convergence seems to be rather irregular both with respect to the
level of electron correlation and basis set size, one should not necessarily
expect more expensive calculations to be much closer to experiment than simpler ones.

For this reason, when possible, experimental data were used to adjust the
\emph{ab initio} potential energy curves, in particular, the experimental
studies by Ram and Bernath  \cite{96RaBexx.ScH,97RaBexx.ScH} where the $v=0$
and sometimes $v=1$ and $v=2$ vibrational states of seven singlet terms ($X\,
\leftexp{1}{\Sigma^+}$, $A\,{}\leftexp{1}{\Delta}$, $B\,{}\leftexp{1}{\Pi}$,
$C\,{}\leftexp{1}{\Sigma^+}$, $D\,{}\leftexp{1}{\Pi}$, $F\,{}\leftexp{1}{\Sigma^-}$
and $G\,{}\leftexp{1}{\Pi}$) were characterised. Of these the $X$, $A$ and $B$
terms dissociate to channel~1 of table~\ref{tbl.Sc.atom.experiment}, $C$ to
channel 7 or perhaps 10, $F$ to channel 6 while for terms $D$ and $G$ channels
3, 6, 7, or 10 are all possible on the basis of symmetry considerations.

Details on the refinement are given in section~\ref{section.linelist};
in the rest of this section we discuss the \emph{ab initio} calculations.

\subsection{Potential energy curves}
Energy curves for the six molecular electronic terms correlating with
the ground atomic states (dissociation channel 1 of
table~\ref{tbl.Sc.atom.experiment})
were computed using CASSCF and internally-contracted MRCI  \cite{Werner1988}
in conjunction with the recent aug-cc-pwCV$n$Z basis sets (awc$n$z for short)
\cite{05BaPexx.ScH,06BaPexx.ScH}.
CASSCF orbitals (state-averaged over all the degenerate components of the six terms considered in this work) were used
as a basis of the MRCI runs.

A four-electron, ten-orbital complete active space
comprising the scandium $4s,3d,4p$ orbitals and the hydrogen $1s$ orbital
(5 active orbitals of a$_1$ symmetry, 2 b$_1$, 2 b$_2$ and 1 a$_2$ in the C$_{2v}$ point
group) was used in the calculations. The outer-core scandium $3s,3p$ orbitals were left doubly occupied at the CASSCF
stage but were correlated at the MRCI one. The inner-core $1s,2s,2p$ orbitals were not
correlated. As discussed by Balabanov and
Peterson \cite{06BaPexx.ScH}, in multireference calculations the late
transition metals Fe-Cu require an active space larger than the full-valence
one, which should include a further set of diffuse $d$ functions.
However this is in not necessary for scandium. Inclusion of the Sc $4p$ orbitals
is not thought to be indispensable for a correct description of bonding but
was found to be necessary in practice to avoid convergence problems
at the CASSCF stage.
All curves were computed in the range 2.0 to 8.5~a$_0$ in steps of 0.05~a$_0$ and
from 9.0 to 13.5~a$_0$ is steps of 0.5~a$_0$, for a total of 141 points.

Our best \emph{ab initio} results are based on MRCI using the awc5z basis set;
computing at this level the energies for all six terms for a single geometry
takes about 4~GB of RAM, 20~GB of disk space and 12 hours on a single core
of an Intel Xeon E5-2670 CPU at 2.60~GHz.
Potential energy curves include a relativistic correction computed as expectation
value of the MVD1 operator.
The Davidson correction was not included in our final \emph{ab initio} curves because,
as already registed for the excitation energies of the
scandium atom (see table~\ref{tbl.Sc.atom.3z}), it does not improve agreement with known
experimental data; furthermore tests (not discussed here in detail)
at the frozen-core / cc-pVDZ level showed that Davidson-corrected energies agreed
worse with full CI than uncorrected MRCI ones.

The \emph{ab initio} energy curves were slightly shifted in energy (i.e., their
$T_e$ were changed) so that they are exactly degenerate upon dissociation; in
our MRCI calculation the exact degeneracy of the terms at $r=+\infty$ is broken
mainly because the energies of (singlet or triplet) $\Sigma^+$ and $\Delta$
terms is computed simultaneously with a two-state calculation, while the
(singlet or triplet) $\Pi$ terms were computed with one-state calculations;
because of the internal contraction approximation used in Molpro in a
multi-state calculation the variational flexibility of the MRCI wave
function is increased and this
results in a small downwards shift in energy. As a consequence at
dissociation the two $\Pi$ terms are about 50~\cm{} higher in energy than the
other terms; furthermore a small breaking of exact degeneracies is expected and
normal also in uncontracted MRCI calculations because of the incomplete
treatment of electron correlation. We therefore thought it was reasonable to
shift all terms to restore the exact degeneracy. The shifts applied to
MRCI/awc5z/MVD1 curves for the $A\,{}\leftexp{1}{\Delta}$,
$B\,{}\leftexp{1}{\Pi}$, $a~\leftexp{3}{\Delta}$, $b\,{}\leftexp{3}{\Pi}$ and
$c\,{}\leftexp{3}{\Sigma^+}$ terms are respectively 1.1, 49.7, 1.5, 49.9 and
2.1~\cm. The ground $X\,{}\leftexp{1}{\Sigma^+}$ term was taken as a reference
and not shifted.

\begin{figure}
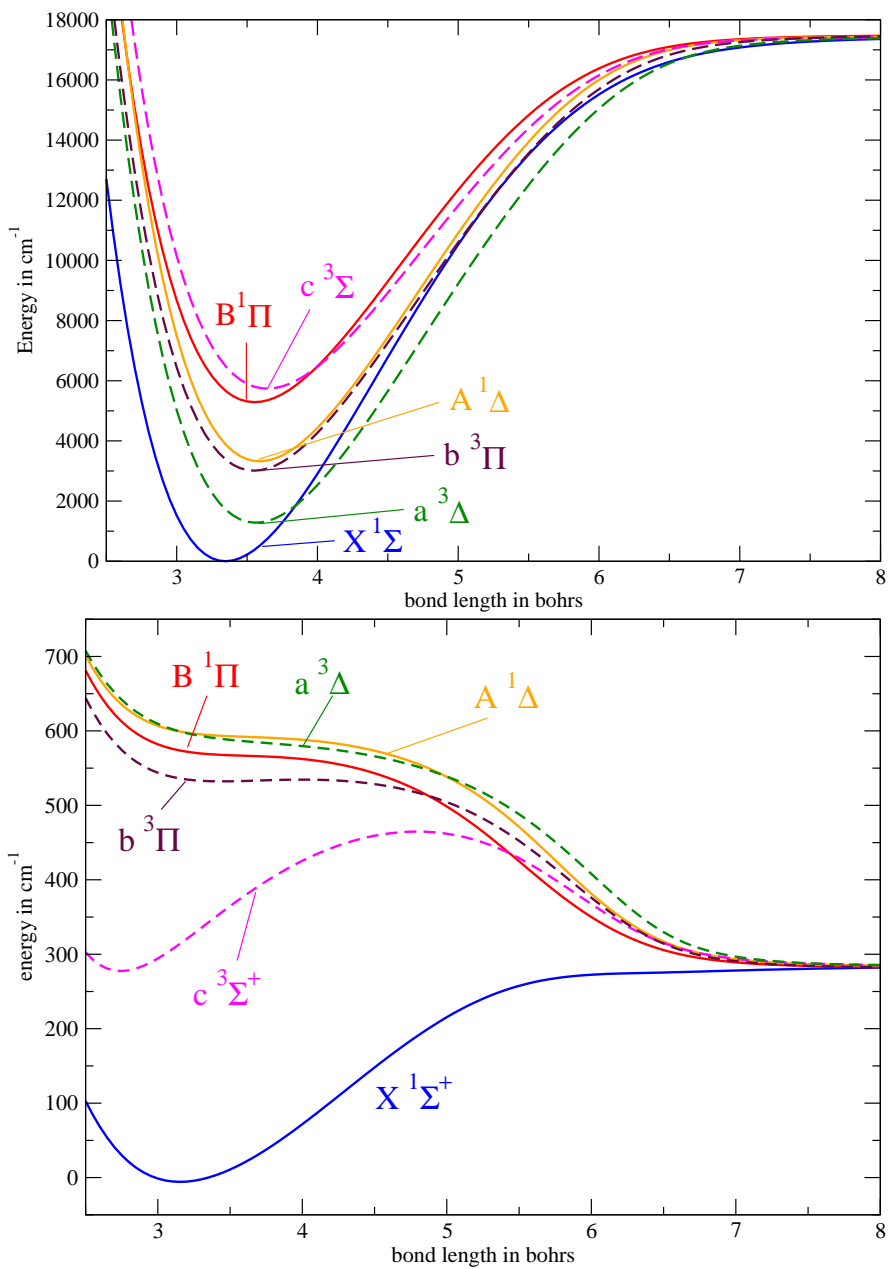

\begin{center}
\includegraphics[angle=0, width=0.85\textwidth]{ScH_awc5z_mrci.eps}
\includegraphics[angle=0, width=0.85\textwidth]{awc5z_mvd1_mrci_3s3p_correlated.eps}
\caption{\emph{Ab initio} potential curves for ScH computed with MRCI and the awc5z basis set and the corresponding relativistic MVD1 correction curves (see text for details). The $3s3p$ orbitals were correlated. \label{energy.curves}}
\end{center}
\end{figure}

Figure~\ref{energy.curves} presents our computed potentials. As it can be seen
the ground $X\,{}\leftexp{1}{\Sigma^+}$ curves are distinct from the other
curves: it has a much shorter equilibrium bond length and its relativistic
correction curve is also very different from the others. This
is a consequence of the
different bonding character of this  $X\,{}\leftexp{1}{\Sigma^+}$ term discussed in section~\ref{sec:intro}.

Equilibrium bond lengths $r_e$, harmonic vibrational frequencies $\omega_e$ and
adiabatic excitation energies $T_e$ are reported in table~\ref{tbl.eq} and
compared with previous theoretical calculations.

Our computed equilibrium bond lengths are in very good agreement (within
0.01~a$_0$) with the recent theoretical values by Hubert \emph{et al}
\cite{13HuOlLo.ScH}.


We compare our \emph{ab initio} and empirically-refined results %
with energy levels reconstructed from the experimental study \cite{97RaBexx.ScH}
in table~\ref{tbl.Obs-Calc}; empirical refinement is discussed in section~\ref{section.linelist}.
We prefer to compare directly with experimental energy levels because
experimentally-deduced values for $T_e$, $r_e$ and $\omega_e$ also include in an effective
way spin-orbit and other coupling effects between different electronic terms.

\begin{table*}
\begin{center}
 \caption{Computed \emph{ab initio} adiabatic electronic excitation energies (in cm$^{-1}$), equilibrium bondlengths (in a$_0$)
 and harmonic frequences (in cm$^{-1}$) for selected ScH electronic terms.} \label{tbl.eq}
\begin{tabular}{l r r r r r r  r r r lll}
\hline
\hline
Term                       & \multicolumn{3}{c}{Anglada et al$^a$}  & \multicolumn{3}{c}{Hubert et al$^b$}                                   &   \multicolumn{3}{c}{This work$^c$}   \\     
\cline{2-4}\cline{5-7}\cline{8-10}\cline{11-13}
                           &   $T_e$   & $r_e$   &  $\omega_e$      & $T_e$                  & $r_e$                 &  $\omega_e$           &      $T_e$ & $r_e$ &  $\omega_e$             \\
$X\,{}\leftexp{1}{\Sigma^+}$  &       0   & 3.41    &  1621            &  0                     &   3.35                &    1611               &       0      &  3.34  &  1587    \\                   
$A\,{}\leftexp{1}{\Delta}$    &    6600   & 3.68    &  1541            &  5362                  &   3.58                &    1439               &    3914      &  3.59  &  1428    \\                   
$B\,{}\leftexp{1}{\Pi}$       &    8400   & 3.64    &  1451            &\multicolumn{1}{c}{---} &\multicolumn{1}{c}{---}&\multicolumn{1}{c}{---}&    5856      &  3.55  &  1380    \\                   

$a\,{}\leftexp{3}{\Delta}$    &    4600   & 3.66    &  1460            &  3660                  &    3.55               &   1450                &   1868     &   3.56   &  1432    \\                 
$b\,{}\leftexp{3}{\Pi}$       &    6200   & 3.64    &  1438            &\multicolumn{1}{c}{---} &\multicolumn{1}{c}{---}&\multicolumn{1}{c}{---}&   3544     &   3.55   &  1406    \\                 
$c\,{}\leftexp{3}{\Sigma^+}$  &    7900   & 3.68    &  1389            &\multicolumn{1}{c}{---} &\multicolumn{1}{c}{---}&\multicolumn{1}{c}{---}&   6122     &   3.63   &  1325    \\                 
\hline
\hline
\end{tabular}
\end{center}
$^a$ Ref. \cite{89AnBrPe.ScH}; values of the $T_e$'s are taken from the column labelled `B(1f)' of table~8, $r_e$'s and $\omega_e$'s from table~10.\\
$^b$ Ref. \cite{13HuOlLo.ScH}, using the data from the column labelled CCSDT12 (Q$\zeta$) of Table V and adding the relativistic corrections in the last column of Table VI.\\
$^c$ Using MRCI/awc5z/MVD1. The $\omega_e^{(i)}$ relative to state $i$ was computed by $\omega_e^{(i)} = \sqrt{ V''(r_e^{(i)}) / \mu }$ while the adiabatic excitation energy $T_e^{(i)}$ was computed as $V_i( r_e^{(i)}) -V_0( r_e^{(0)})$, where $V_0$ is the potential for the $\leftexp{1}{\Sigma^+}$ ground state.\\
\end{table*}

\begin{table*}
\footnotesize
\begin{center}
\caption{Selected energy levels for the lowest-energy singlet terms of ScH in \cm. $J$ is the total angular momentum (neglecting nuclear spin), $v$ is the  vibrational quantum number, $p$ the parity; `Obs.' are the values derived using the spectroscopic constants reported by Ram and Bernath \cite{97RaBexx.ScH} and the program PGOPHER~\cite{PGOPHER}; `A'
and `R' are the term values calculated with {\sc Duo} using the {\it ab initio} (MRCI/awc5z/MVD1) and {\it refined} curves as described  in the text. Calculations included spin-orbit and other couplings between electronic terms.
\label{tbl.Obs-Calc} }
\begin{tabular}{rcrrrrr}
\hline
\hline
$J$     &  $p$          &   \multicolumn{3}{c}{energies}      &  \multicolumn{2}{c}{obs -- calc} \\
        &               &  \multicolumn{1}{c}{obs}&  \multicolumn{1}{c}{\emph{ab initio}} & \multicolumn{1}{c}{\emph{refined}}   &  \multicolumn{1}{c}{\emph{ab initio}} & \multicolumn{1}{c}{\emph{refined}}   \\
\mbox{}\\
\multicolumn{7}{c}{$X\,{}^1\Sigma^+$, $v=0$}\\
    0  &    +   &         0.00  &        0.00  &      0.00 &$       0.00 $&$      0.00 $\\
    1  &    -   &        10.73  &       10.73  &     10.73 &$       0.00 $&$      0.00 $\\
   10  &    +   &       586.89  &      586.88  &    586.93 &$      -0.04 $&$     -0.05 $\\
\multicolumn{7}{c}{$X\,{}^1\Sigma^+$, $v=1$}\\
    0  &    +   &      1546.97  &     1538.53  &   1547.10 &$       8.44 $&$     -0.12 $\\
    1  &    -   &      1557.45  &     1549.00  &   1557.54 &$       8.45 $&$     -0.09 $\\
   10  &    +   &      2120.06  &     2111.48  &   2118.31 &$       8.58 $&$      1.76 $\\
\mbox{}\\
\multicolumn{7}{c}{A~$^1\Delta$, $v=0$}\\
    2  &    +   &      4213.36  &     3830.27  &   4213.35 &$     383.09 $&$      0.01 $\\
    3  &    +   &      4241.34  &     3858.12  &   4241.36 &$     383.21 $&$     -0.03 $\\
   10  &    +   &      4696.30  &     4311.04  &   4696.64 &$     385.25 $&$     -0.34 $\\
   10  &    -   &      4696.30  &     4311.01  &   4696.61 &$     385.28 $&$     -0.31 $\\
\mbox{}\\
\multicolumn{7}{c}{B~$^1\Pi$, $v=0$}\\
    1  &    +   &      5413.98  &     5704.18  &   5413.94 &$    -290.19 $&$      0.04 $\\
    2  &    +   &      5433.77  &     5723.68  &   5433.79 &$    -289.91 $&$     -0.02 $\\
   10  &    +   &      5943.02  &     6226.10  &   5939.67 &$    -283.07 $&$      3.35 $\\
   10  &    -   &      5939.42  &     6222.43  &   5936.36 &$    -283.01 $&$      3.06 $\\
\multicolumn{7}{c}{B~$^1\Pi$, $v=1$}\\
    1  &    +   &      6776.75  &     7037.67  &   6776.70 &$    -260.92 $&$      0.05 $\\
    2  &    +   &      6795.96  &     7056.69  &   6796.35 &$    -260.73 $&$     -0.39 $\\
   10  &    +   &      7290.12  &     7546.50  &   7290.43 &$    -256.38 $&$     -0.31 $\\
   10  &    -   &      7286.57  &     7542.84  &   7286.95 &$    -256.28 $&$     -0.38 $\\
\multicolumn{7}{c}{B~$^1\Pi$, $v=2$}\\
    1  &    +   &      8092.50  &     8323.48  &   8092.51 &$    -230.98 $&$     -0.01 $\\
    2  &    +   &      8111.15  &     8342.00  &   8111.07 &$    -230.85 $&$      0.08 $\\
   10  &    +   &      8590.66  &     8818.77  &   8589.10 &$    -228.10 $&$      1.56 $\\
   10  &    -   &      8587.14  &     8815.10  &   8585.06 &$    -227.96 $&$      2.08 $\\
\hline
\end{tabular}
\end{center}
\end{table*}

As discussed above, we expect our computed adiabatic excitation energies $T_e$
to have errors of several hundreds or perhaps a few thousands of \cm{} and
therefore they cannot be considered very accurate. On the other hand the
equilibrium bond lengths and the shape of the potentials should be reasonably
accurate, see also table~\ref{tbl.Obs-Calc}, where the accuracy of the
potential energy curves of the singlet states can be assessed by comparing to
the vibrational energy separations within each state.
The rotational intervals between the two lowest-$J$ $v=0$ energy levels are reproduced
by our \emph{ab initio} curves with errors of 0.00, 0.12 and 0.28~\cm\ for the
$X$, $A$ and $B$ term respectively. Errors in vibrational transitions $v=0$ to $v=0$
are 8.44~\cm\ for the $X$ term and 29.27~\cm\ for the $B$ term.

For the ground X term only,  we considered the error of the coupled-cluster 
based potential curves. The absolute errors for the $v=0$ to $v=1$ 
transition wavenumber using CCSD, CCSD(T), CCSDT and CCSDT(Q) are, 
respectively, 54.73, 24.45, 6.75 and 1.15~\cm\ with the awc5z basis set, 
the DKH Hamiltonian and keeping all the coupling terms computed with MRCI or CASSCF; the CCSDT and CCSDT(Q) curves were obtained in the basis set formed 
by wc3z for Sc and 2z for hydrogen, and added as a correction to the CCSD(T) curve.
These result indicate that our MRCI curves are, close to equilibrium, similar in quality to CCSDT and that quadruple excitations must be accounted for to obtain accuracies of the order of 1~\cm.

\subsection{Dissociation energy}
The dissociation energy $D_0$ of ScH is related
to the potential well depth $D_e$ by
\begin{equation}
D_0  = D_e - E_\mathrm{ZPE}
\end{equation}
where $E_\mathrm{ZPE}=787$~\cm{} is the zero-point rotational-vibrational
energy and the quoted value was computed using out MRCI/awc5z/MVD1 PEC. The
potential well depth $D_e$ can be decomposed into three contributions: a main
non-relativistic one, a spin-independent (scalar) relativistic contribution and
a spin-dependent contribution due to spin-orbit:
\begin{equation}
D_e  = D_e^\mathrm{NR} + D_e^\mathrm{R} + D_e^\mathrm{SO}
\end{equation}
The spin-orbit contribution $D_e^\mathrm{SO}$ is due to the energy lowering
of the scandium atom $^2\mathrm{D}_{3/2}$ level with respect
to the $^2\mathrm{D}$ term 
and has a value $D_e^\mathrm{SO} = -3\xi/2 = -101$~\cm, where
$\xi = 67.3$~\cm{} is the atomic spin-orbit contant for the $^2\mathrm{D}$ term
(see table~\ref{tbl.Sc.atom.experiment}).

The (spin-orbit free) potential well depth $D_e^\mathrm{NR}$ computed with
MRCI/awc5z is 17459~\cm;
the Davidson correction gives a rather large shift of +1355~\cm,
leading for MRCI+Q/awc5z to a value $D_e^\mathrm{NR}$=18814~\cm.

With a view to ascertaining the quality of our \emph{ab initio} curve close
to dissociation we computed an accurate value for $D_e^\mathrm{NR}$ using high-order
coupled cluster and the program MRCC \cite{mrcc}.

Using the awc5z basis set and correlating the outer-core $3s3p$ orbitals
gives for $D_e^\mathrm{NR}=17694$~\cm{} using CCSD and 18547~\cm{} using CCSD(T).
The effect of full triples (T)$\rightarrow$T
was evaluated in a basis set formed by the wc3z for scandium and 2z for hydrogen,
giving a shift of +163~\cm. The effect of quadruple excitations was evaluated
in an even smaller basis set constucted complementing the 2z one for hydrogen and scandium
with the core-correlation functions taken from the wc3z basis set and dropping the $g$ functions;
the computed shift T$\rightarrow$Q is +48~\cm; our best awc5z
(frozen inner-core) coupled cluster value is $D_e^\mathrm{NR}=18547+163+48 = 18758$~\cm.
The coupled cluster value therefore strongly supports the Davidson-corrected
value for $D_e^\mathrm{NR}$ rather than the uncorrected MRCI one.

We also considered the contribution to $D_e^\mathrm{NR}$  due to correlation of the
inner-core $2s2p$ orbitals; this effect gives a contribution of +62~\cm\ using CCSD/awc5z and
+104~\cm{} using CCSD(T)/awc5z.

The magnitude of basis set incompleteness was estimated by looking at the difference
between the awc5z values and the basis set extrapolated ones
using the awc4z and awc5z basis sets;
Basis set extrapolation of the awc5z values gives a very small contribution, namely
$-3$~\cm{} for MRCI/awc5z and +6~\cm{} for CCSD(T)/awc5z.

Finally, our best coupled-cluster-based value for the non-relativistic part of
the potential well depth is $D_e^\mathrm{NR} =18758+104+6=18868(50)$~\cm, where the
given uncertainty was assigned by halving of the sum of the quadruples correction,
the difference between the CCSD and CCSD(T) inner-core corrections and the basis set extrapolation
contribution.

We now consider the scalar relativistic contribution $D_e^\mathrm{R}$.
The MVD1/awc5z value is $D_e^\mathrm{R}=+285$~\cm; using the DKH Hamiltonian (truncated to
fourth order) and the awc4z-DK basis gives a contribution +311~\cm{} using MRCI,
+291~\cm{} using MRCI+Q and +307~\cm{} using CCSD(T).

Taking the DKH value $D_e^\mathrm{R} = 307$~\cm{} (although there is no conclusive
argument to favour it instead of the MVD1 one) we arrive at a final value for the potential
well depth $D_e = 18868+307-101=19074(60)$~\cm\ and to a
dissociation energy $D_0 = 19074 -787 = 18287(60)$~\cm,
where the given uncertainty was increased to reflect the uncertainty on
the relativistic correction.

Kant and Moon \cite{81KaMoxx.ScH} reported long ago an experimental value for
the potential well depth $D_e = 16613 \pm 700$ \cm\ and Koseki {\it et al}
\cite{04KoIsFe.ScH} gave a survey of calculated values of $D_e$ which have a
large spread of about 3000~\cm\ around the value quoted. Our computed value is
larger than the experimental one by about 2500~\cm\ and in disagreement with it
by more than three times its uncertainty bar.

Finally, we report some run-times for our coupled cluster results.
The CCSDT run using the wc2z/2z basis set took for ScH 8.3 hours of CPU time on a
12-core Xeon X5660 at 2.80GHz machine (1.3 hours real time). The CCSDTQ run for ScH
in the 2z+wC3z basis set took 10.6 days of CPU time (1.6 days real time) and 5~GB of
RAM on the same machine. A single CCSD(T)/awc5z run takes about 10 minutes on a single core
of the same machine.

\subsection{Dipole moment curves}
While   potential energy curves can be refined semiempirically from (even
limited) experimental data, one has normally to rely on computed, \emph{ab
initio} dipole moment curves \cite{jt156}. With a view to computing accurate line
intensities it is therefore important to produce dipole moment curves as
accurate as possible.

Le and Steimle \cite{11LeStxx.ScH} reported for the ground $X\,{}^1 \Sigma^+$
term an experimental equilibrium dipole $\mu = 1.74$(0.15)~D using optical
Stark spectroscopy; for our work we choose the $z$ axis such that a negative
dipole corresponds to Sc$^+$H$^-$ polarity, so we reverse their value to $\mu
=-1.74$(0.15)~D.

We were able to produce a very accurate value for the equilibrium dipole of the
ground state term using coupled cluster and the energy-derivative (ED) method
\cite{jt475} ($\lambda=\pm 10^{-4}$~au). As we are dealing with a closed-shell
electronic state near equilibrium coupled cluster converges quickly with
respect to the level of excitations included. High-order coupled cluster
calculations used the program MRCC \cite{mrcc}. Results are collected in
table~\ref{tab:cc.eq.dipole}; our best theoretical value is $\mu_e =-1.72(2)$~D
and is in full agreement with the experimental value of Le and Steimle
\cite{11LeStxx.ScH}.

\begin{table*}
\begin{center}
\caption{Equilibrium dipole of the ground state $X\,{}^{1}{\Sigma^+}$ term using coupled cluster theory.
Dipoles were computed at $r=3.35$~a$_0$ using the energy-derivative method and $\lambda=\pm10^{-4}$~au.
A part from the line labelled `D', all calculations correlated the outer-core $3s3p$ orbitals but the kept the inner core $1s2s2p$ uncorrelated.
Dipoles are in debyes. The experimental value is -1.74$(0.15)$~D \cite{11LeStxx.ScH}.
\label{tab:cc.eq.dipole}}
\begin{tabular}{c l l l r}
\hline
\hline
\multicolumn{1}{c}{label}  & \multicolumn{1}{c}{method} & \multicolumn{1}{c}{basis set} & \multicolumn{1}{c}{value} \\
                           &RHF          & awc3z         & $-$1.436\\
                           &CCSD         & awc3z         & $-$1.668\\
                           &CCSD(T)      & awc3z         & $-$1.640\\
                           &CCSD(T)      & awc4z         & $-$1.686\\
                           &CCSD(T)      & awc5z         & $-$1.702\\
A                          &CCSD(T)      & awc[345]z$^a$ & $-$1.719\\
\mbox{}\\
&\multicolumn{3}{c}{higher order correlation}\\
B                          & (T)  $\rightarrow$ T        & wc3z/2z$^b$     & $-$0.002\\
C                          &  T   $\rightarrow$ T(Q)     & wc3z/2z$^b$     & $-$0.009\\
D                          & T(Q) $\rightarrow$ Q        & 2z+wC(3z)$^c$   & $-$0.005\\
\mbox{}\\
&\multicolumn{3}{c}{inner-core correlation$^d$}\\
E                          & CCSD(T)                     & awc3z           & $-$0.011\\
\mbox{}\\
&\multicolumn{3}{c}{relativistic$^e$}\\
F                          & CCSD(T)                     & awc4z           & +0.078\\
\mbox{}\\
&\multicolumn{3}{c}{vibrational averaging$^f$}\\
G                          &                             &                 & $-$0.054\\
\hline
\hline
A + B +C +D +E+F+G         & best \emph{ab initio}$^g$   &                 & $-$1.72(2)\\
\hline
\hline
\end{tabular}
\end{center}
$^a$ Basis-set extrapolated value using a $\mu_n = \mu_e + A / n^3$ formula.\\
$^b$ Correction due to full triples and perturbative quadruple excitations using the cc-pVDZ basis set for hydrogen and cc-pwCVTZ for scandium.\\
$^c$ Correction due to full quadruple excitations using the cc-pVDZ basis set for hydrogen and the cc-pVDZ complemented with the core-valence correlation functions from the cc-pwCVTZ ($g$ functions excluded) for scandium.\\
$^d$ Correction due to correlation of the $2s2p$ orbitals. The innermost $1s$ orbital was not correlated.\\
$^e$ Correction due to scalar-relativistic effects computed as difference of CCSD(T)/DKH4/awc5z-dk and CCSD(T)/awc5z dipoles.\\
$^f$ Vibrational averaging computed as $\langle 0 | \mu(r) | 0 \rangle - \mu(r=3.35 \mathrm{a}_0)$, where $|0\rangle$ is the $J=0, v=0$ vibrational
ground state (obtained using the \emph{ab initio} MRCI/awc5z PEC) and $\mu(r)$ is the MRCI/energy-derivative/awc5z dipole moment curve.\\
$^g$ The estimated uncertainty in the teoretical dipole is mostly due to residual basis set incompleteness error and incomplete treatment of higher-order correlation effects.
\end{table*}

\begin{figure}
\begin{center}
\includegraphics[angle=0, width=0.85\textwidth]{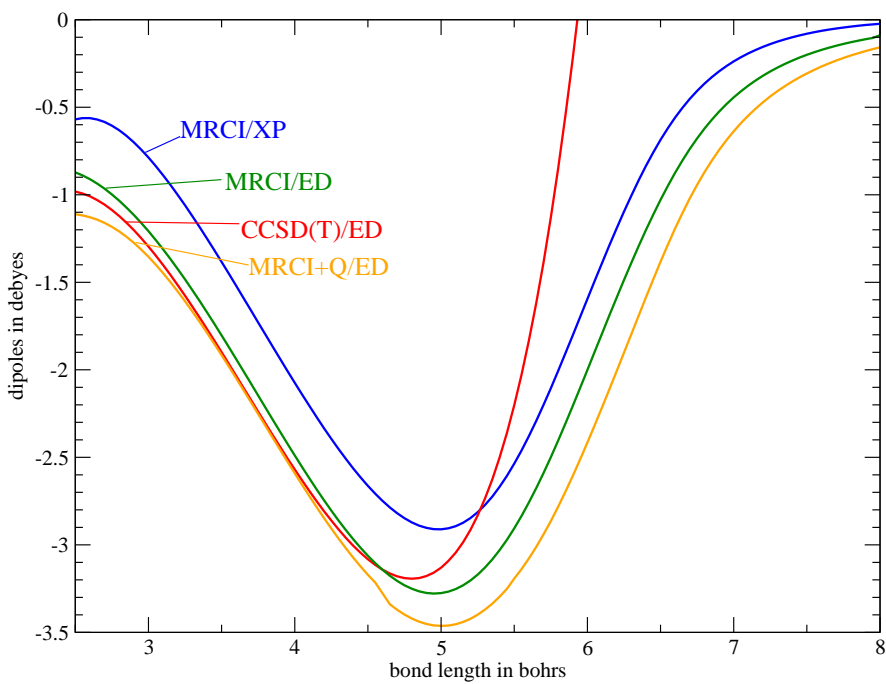}
\caption{\emph{Ab initio} dipole curves for the ground  $X\,{}^{1}{\Sigma^+}$ term computed in the awc5z basis set and various methods (see text).  \label{fig.1S.dipoles}}
\end{center}
\end{figure}

MRCI, on the other hand, has difficulties in reproducing the correct value for
the dipole. Apart from the basis set size and whether or not core orbitals are
correlated, we considered three further factors affecting MRCI dipoles, namely:
\emph{i)} whether dipoles are computed by expectation value (XP) or energy
derivative (ED) \cite{jt475}; \emph{ii)} whether at the CASSCF step orbitals
are obtained by state averaging (state-averaged orbitals, SAO) or are
specifically optimised for for the $X\,{}^{1}{\Sigma^+}$ electronic term
(state-specific orbitals, SSO); and, \emph{iii)}, the effect of using
Davidson-corrected energies instead of MRCI ones. We compared dipoles obtained
with MRCI in the awc3z basis set with the very accurate (non-relativistic,
$3s3p$ correlated) value obtained with coupled cluster, which gives in this
basis set the value $\mu = 1.66$~D (see table~\ref{tab:cc.eq.dipole},
CCSD(T)/awc3z value + lines B,C,D). Results are collected in
table~\ref{tab:mrci.eq.dipole}. Figure~\ref{fig.1S.dipoles} shows the
dipole moment curves for the X ground term computed with CCSD(T), MRCI/XP,
MRCI/ED and MRCI+Q/ED; note that the CCSD(T) curve become unphysical
at large bondlengths.

\begin{table*}
\begin{center}
\caption{Values and errors in computed MRCI dipoles at $r=3.35$~a$_0$ for the ground state $X\,{}^{1}{\Sigma^+}$ term. The acronym SAO and SSO stand for state-averaged orbitals and state-specific orbitals, respectively. The last two columns report differences with the accurate value $\mu = 1.66$~D obtained with coupled cluster (see text). Dipoles are in debyes. \label{tab:mrci.eq.dipole}}
\begin{tabular}{l r r r r}
\hline
\hline
\multicolumn{1}{c}{method$^a$}& \multicolumn{2}{c}{values} & \multicolumn{2}{c}{values -- exact} \\
                              &        SAO   &  SSO           &   SAO   &   SSO \\
CASSCF/XP                     &      -0.99  &  -1.32          &   0.67  &  0.34 \\
MRCI/XP                       &      -1.19  &  -1.35          &   0.47  &  0.31 \\
MRCI/ED                       &      -1.56  &  -1.66          &   0.08  &  0.00 \\
MRCI+Q/ED                     &      -1.68  &  -1.71          &  -0.02  & -0.05 \\
\hline
\hline
\end{tabular}
\end{center}
$^a$ XP or ED specifies whether dipole were computed as expectation value or energy derivatives (field strength $\lambda = \pm 10^{-4}$~au). \\
\end{table*}

As one can see in table~\ref{tab:mrci.eq.dipole} both CASSCF and MRCI/XP
equilibrium dipoles are too small in magnitude by about 0.3--0.5~D, a
considerable amount; using state-specific orbitals reduces the error to about
0.3~D, indicating that the CASSCF and MRCI wave functions are quite far from
the exact, full CI one (which is independent on the choice of the orbitals).
Computing the dipole by energy derivative greatly reduces the error in the
MRCI dipoles, bringing them in much closer agreement with the coupled cluster
value. Using Davidson-corrected energies (MRCI+Q) introduces a shift of about
0.12~D and brings the dipole curve near equilibrium even closer to the coupled
cluster one (see also fig.~\ref{fig.1S.dipoles}). The Davidson-corrected dipole
curve (`relaxed' reference energies were used) has a small jump discontinuity
(0.02~D of magnitude) at $r=4.6$~a$_0$; on the other hand the MRCI/ED curve is
perfectly smooth for all bond lengths. In conclusion the major factor affecting
MRCI dipoles is whether the XP or ED technique is used, with ED producing
considerably better dipoles. Using SSO instead of SSA helps somewhat but is of
secondary importance.

It is also worth noting that, although the absolute value of MRCI/XP
dipoles is considerably off, this quantity only affects line
intensities for pure rotational transitions. Intensities of
vibrational transition within an electronic term depend on the shape
of the dipole function, and may be given quite accurately even by
MRCI/XP.

\begin{figure}
\begin{center}
\includegraphics[angle=0, width=0.85\textwidth]{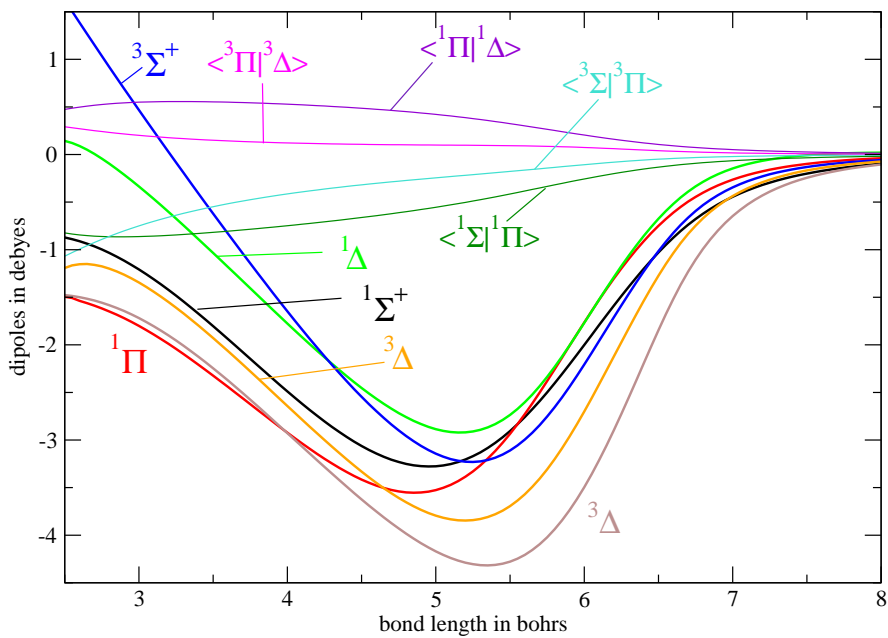}
\caption{Diagonal (in bold) and off-diagonal dipole moment curves for ScH computed with with MRCI and the awc5z basis set. \label{dipole.curves}}
\end{center}
\end{figure}

Finally, we decided to use awc5z/MRCI/ED for all diagonal dipole curves in virtue of their smoothness,
although using MRCI+Q dipoles may lead to a slight improvement. 
Off-diagonal dipoles were computed as expectation values of the awc5z/MRCI wave
functions; although it is possible to compute off-diagonal dipoles using an
energy-derivative technique \cite{98AdZaSt.method}, this route was not pursued
at this time. Figure~\ref{dipole.curves} shows the diagonal and off-diagonal
dipole moment curves for the various electronic.

\subsection{Spin-orbit and other coupling curves}
We computed spin-orbit couplings and couplings of the angular momentum operators $\hat{L}_x$ and $\hat{L}_y$
using the CASSCF or MRCI wave functions.

\begin{figure}
\begin{center}
\includegraphics[angle=0, width=0.85\textwidth]{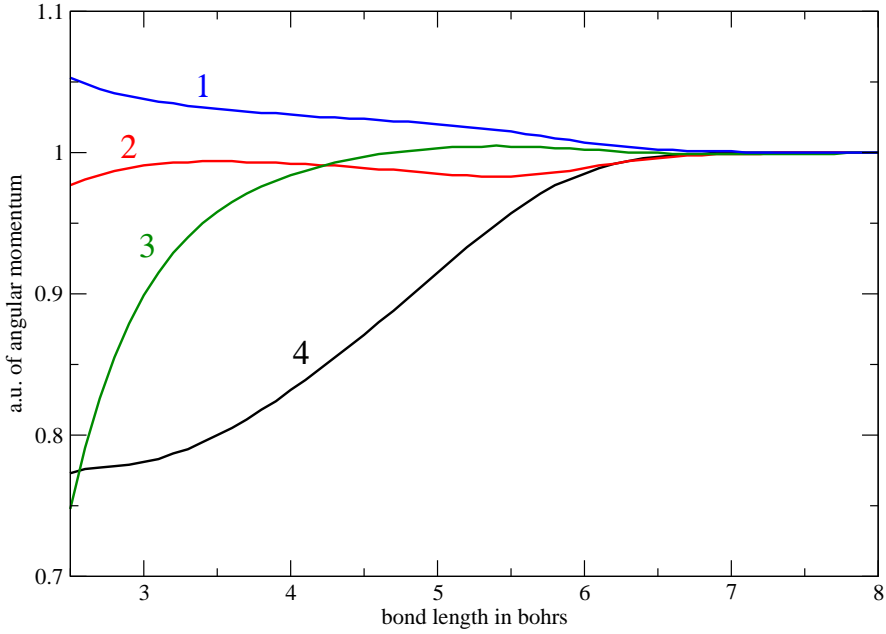}
\caption{Matrix elements of the $\hat{L}_x$ and $\hat{L}_y$ operators for ScH computed with CASSCF and the awc3z basis set. Specifically, the curve labelled `1' is $\langle ^3 \Delta_{x^2-y^2} |\hat{L}_x | ^3 \Pi_y \rangle /i$; curve `2' is $\langle ^1 \Delta_{x^2 -y^2} |\hat{L}_x | ^1 \Pi_y \rangle /(-i)$; curve `3' is  $\langle ^3 \Sigma^+ |\hat{L}_y | ^3 \Pi_x \rangle /(\sqrt{3} i)$ ; curve `4' is $\langle ^1 \Sigma^+ |\hat{L}_y | ^1 \Pi_x \rangle /(\sqrt{3} i)$. The phases of the electronic wave functions are chosen such as $\langle ^1 \Pi_x |\hat{L}_z | ^1 \Pi_y \rangle=i$, $\langle ^3 \Pi_x |\hat{L}_z | ^3 \Pi_y \rangle=i$, $\langle ^1\ \Delta_{x^2-y^2} |\hat{L}_z | ^1\Delta_{xy} \rangle=-2i$ and $\langle ^3\ \Delta_{x^2-y^2} |\hat{L}_z | ^3\Delta_{xy} \rangle=2i$.  \label{LxLy.curves}}
\end{center}
\end{figure}

\begin{figure}
\begin{center}
\includegraphics[angle=0, width=0.85\textwidth]{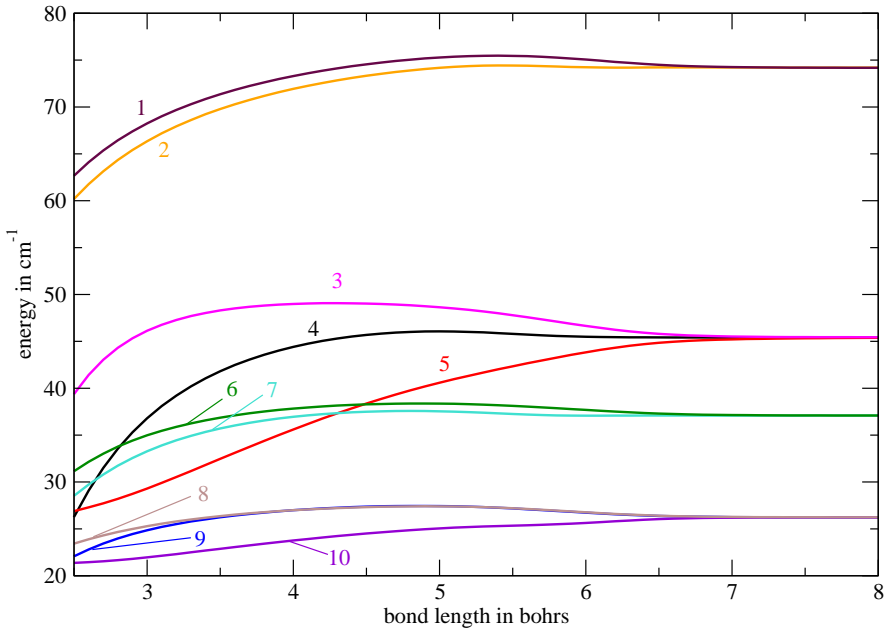}
\caption{Spin-orbit coupling matrix elements for ScH  computed with CASSCF and the awc3z basis set. Specifically: the curve labelled `1' is $\langle ^3 \Delta_{x^2-y^2}, \Sigma=1 |\hat{H}_\mathrm{SO} | ^3 \Delta_{xy}, \Sigma=1 \rangle /(-i)$; curve `2'  is $\langle ^1 \Delta_{xy} |\hat{H}_\mathrm{SO} | ^3 \Delta_{xx-yy} , \Sigma=0\rangle /(-i)$; curve `3'  is $\langle ^3 \Sigma^+,    \Sigma=   0 |\hat{H}_\mathrm{SO} | ^3 \Pi_y          , \Sigma=1\rangle /(-i)$; curve `4'  is $\langle ^1 \Pi_y  |\hat{H}_\mathrm{SO} | ^3 \Sigma^+       , \Sigma=1\rangle /i$; curve `5'  is $\langle ^1 \Sigma^+    |\hat{H}_\mathrm{SO} | ^3 \Pi_y          , \Sigma=1\rangle /i$; curve `6'  is $\langle ^3 \Pi_x    ,   \Sigma=   1 |\hat{H}_\mathrm{SO} | ^3 \Pi_y          , \Sigma=1\rangle /(-i)$; curve `7'  is $\langle ^1 \Pi_x     |\hat{H}_\mathrm{SO} | ^3 \Pi_y          , \Sigma=0\rangle /i$; curve `8'  is $\langle ^3 \Pi_x    ,   \Sigma=   0 |\hat{H}_\mathrm{SO} | ^3 \Delta_{xy}    , \Sigma=1\rangle /i$; curve `9'  is $\langle ^1 \Delta_{xy}|\hat{H}_\mathrm{SO} | ^3 \Pi_x          , \Sigma=1\rangle /(-i)$; curve `10' is $\langle ^1 \Pi_y      |\hat{H}_\mathrm{SO} | ^3 \Delta_{xx-yy} , \Sigma=1\rangle /(-i)$. The phases of the electronic wave functions are chosen such as $\langle ^1 \Pi_x |\hat{L}_z | ^1 \Pi_y \rangle=i$, $\langle ^3 \Pi_x |\hat{L}_z | ^3 \Pi_y \rangle=i$, $\langle ^1\ \Delta_{x^2-y^2} |\hat{L}_z | ^1\Delta_{xy} \rangle=-2i$ and $\langle ^3\ \Delta_{x^2-y^2} |\hat{L}_z | ^3\Delta_{xy} \rangle=2i$.\label{SO.curves}}
\end{center}
\end{figure}

Figure \ref{LxLy.curves} shows matrix elements of the $L_x$ and $L_y$ operator, obtained at the CASSCF/awc3z level;
these couplings enter in the $L$-uncoupling and spin-electronic terms of the
rotational Hamiltonian \cite{86LeFexx} and are responsible for $\Lambda$-doubling.
Finally, fig.~\ref{SO.curves} reports the 10 symmetry-independent spin-orbit coupling curves obtained at the
CASSCF/awc3z level. Care was taken to ensure that the coupling curves and dipoles are both smooth and phase
corrected, something that is by no means standard in the literature \cite{jt573}.

\section{Line list}\label{section.linelist}
The potential energy, dipole and coupling curves were then used with the
in-house program {\sc Duo} to produce a line list for $^{45}$ScH. {\sc Duo}
solves in an essentially exact way the rotational-vibrational-electronic
problem for multiple interacting energy curves for diatomic molecules and is
described in detail elsewhere \cite{jt606,jt589,jt598}. The line list can be
obtained from \textit{www.exomol.com}, while all the curves used to produce it
are made available as supplementary material. In all nuclear-motion
calculations we used the atomic masses $m_\mathrm{H}=1.0078250321$~Da and
$m_\mathrm{Sc}=44.9559100$~Da, which give for ScH a reduced mass $\mu =
(m_\mathrm{H}^{-1} +   m_\mathrm{Sc}^{-1})^{-1} =0.985726930$~Da =
1796.87027~$m_e$.

The \emph{ab initio} potential energy curves of the singlet terms were adjusted
by fitting to the energy term values ($J\le 12$) derived from the experimental
spectroscopic constants reported by Ram and Bernarth \cite{97RaBexx.ScH} which
cover vibrational excitations with $v=0,1$ ($X$), $v=0$ ($A$) and $v=0,1,2$
($B$) only. This was not possible for the triplet states (see
table~\ref{tbl.eq}). We also decided not to use in the adjustments the very
recent experimental data on the $а{}\leftexp{3}{\Delta}$ electronic state
\cite{14MoBhNa.ScH} since it lead to an equilibrium bond length
$r_e=3.94$~a$_0$ which differs too substantially from theory to be safely
trusted, see table~\ref{tbl.eq}.

In the refined curves the dissociation energy was fixed to the $D_e$ value by
Koseki {\it et al} \cite{81KaMoxx.ScH}. The triplet curves were also 
scaled.
to dissociate to the same value of $D_e$. All refined curves are given as
supplementary material to the paper together with the {\it ab initio} curves.
The triplet electronic states appear to be in the strong resonance with the
rovibronic states from $B~\leftexp{1}{\Pi}$  and prevented an accurate fit to
the $B$-state energies, especially for $J>12$. It should be noted however that
the spectroscopic constants from Ref.~\cite{97RaBexx.ScH} were derived using in
the absence of the of interaction with the triplet states, and thus can also
contain artifacts.

We then used the program {\sc Duo} \cite{jt606} to solve the coupled
Schr\"{o}dinger equations
to compute the rovibronic energies of ScH up to the dissociation. 
In particular, we obtained for $^{45}$ScH a zero-point-energy of 799.6~\cm. The
highest value the total angular momentum $J$ can assume for bound states is
found to be $J=59$.


The corresponding rovibronic eigenfunctions were combined with the \emph{ab
initio} dipole moment curves to produce Einstein~A coefficients for all
transitions with line positions up to $D_0$. The Einstein~A coefficients
together with the rovibrational energies supplemented by the total degeneracies
and quantum numbers make up the line list.

{\sc Duo} calculations consist of two steps: in the first step we used a grid
of 501 points to solve six separate vibrational Schr\"{o}dinger equations for
each electronic state, using as potential the \emph{ab initio} potential curves
shown in Fig.~\ref{energy.curves} or the empirically adjusted curves described
above. We then selected 40 lowest-energy eigenfunctions from each set; the
union of these $40\times 6 = 240$ functions constitutes our vibrational basis
set $|{\rm state},v\rangle $, where $v$ is the vibrational quantum number and
`state' is the label identifying the electronic state. In the second step of
the calculation we build a basis set of Hund's case~a functions of the type
\begin{equation}\label{e:basis}
 |{\rm state}, \Lambda, S, \Sigma, v \rangle  =  | {\rm state}, \Lambda, S, \Sigma,  \rangle | J,\Omega,M \rangle |{\rm state},v\rangle,
\end{equation}
where $ | {\rm state}, \Lambda, S, \Sigma  \rangle$ is the electronic function,
$| J, \Omega,M \rangle$ is the rotational function and $|{\rm state},v\rangle $ is
one of the vibrational functions; $\Lambda$, $\Sigma$, and $\Omega$ are the $z$ axis
projections of the electronic, spin and total angular momenta, respectively,
and $\Omega= \Lambda+\Sigma$; $M$ is the projection of the total angular
momentum along the laboratory axis $Z$. The full, coupled problem is then solved
by exact diagonalization in the chosen basis set.

In order to guarantee that all phases of the \emph{ab initio} couplings as well
as transition dipole moments are consistent, we used the matrix elements of the
$\hat{L}_z$ operator between the corresponding degenerate $\Pi$ and $\Delta$
components as provided by Molpro. These matrix elements were then used to
transform the matrix elements of all coupling to the representation where
$\hat{L}_z$ is diagonal, which is used in {\sc Duo} according with
eq.~(\ref{e:basis}). The phases are chosen such that all matrix elements are
positive.

Our  $^{45}$ScH line list  contains 1~152~827 transitions and is given in the
ExoMol format \cite{jt548} consisting of two files, an energy file and a
transition file. This is based on a method originally developed for the BT2
line list \citep{jt378}. Extracts for the line lists are given in
Tables~\ref{t:Energy-file} and \ref{t:transition-file}. Using all energies of
$^{45}$ScH we computed its partition function for temperatures up to 5000~K.
The line lists and partition function together with auxiliary data including
the potential energy, spin-orbit, electronic angular momentum, and dipole
moment curves, as well as the absorption spectrum given in cross section format
\citep{jt542}, can all be obtained from the ExoMol website at
\textit{www.exomol.com}.

\begin{figure}
\begin{center}
\includegraphics[angle=0, width=0.85\textwidth]{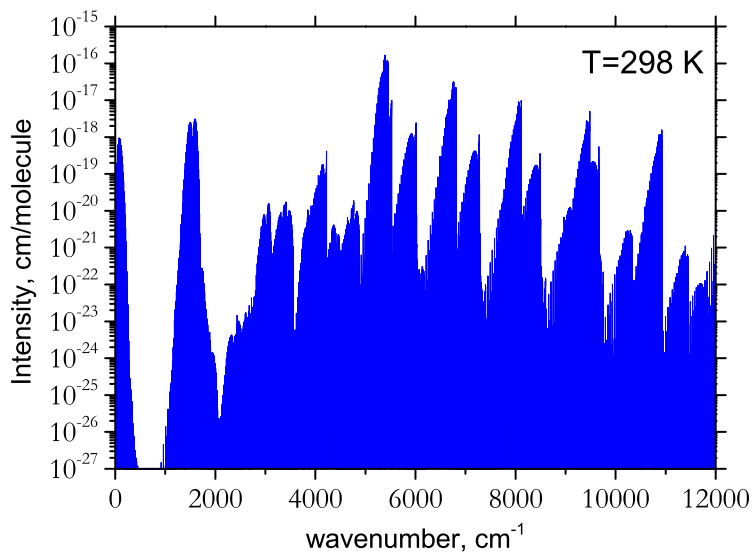}
\caption{Absorption cross-sections of ScH at $T$=298~K. \label{f:298K}}
\end{center}
\end{figure}

\begin{figure}
\begin{center}
\includegraphics[angle=0, width=0.85\textwidth]{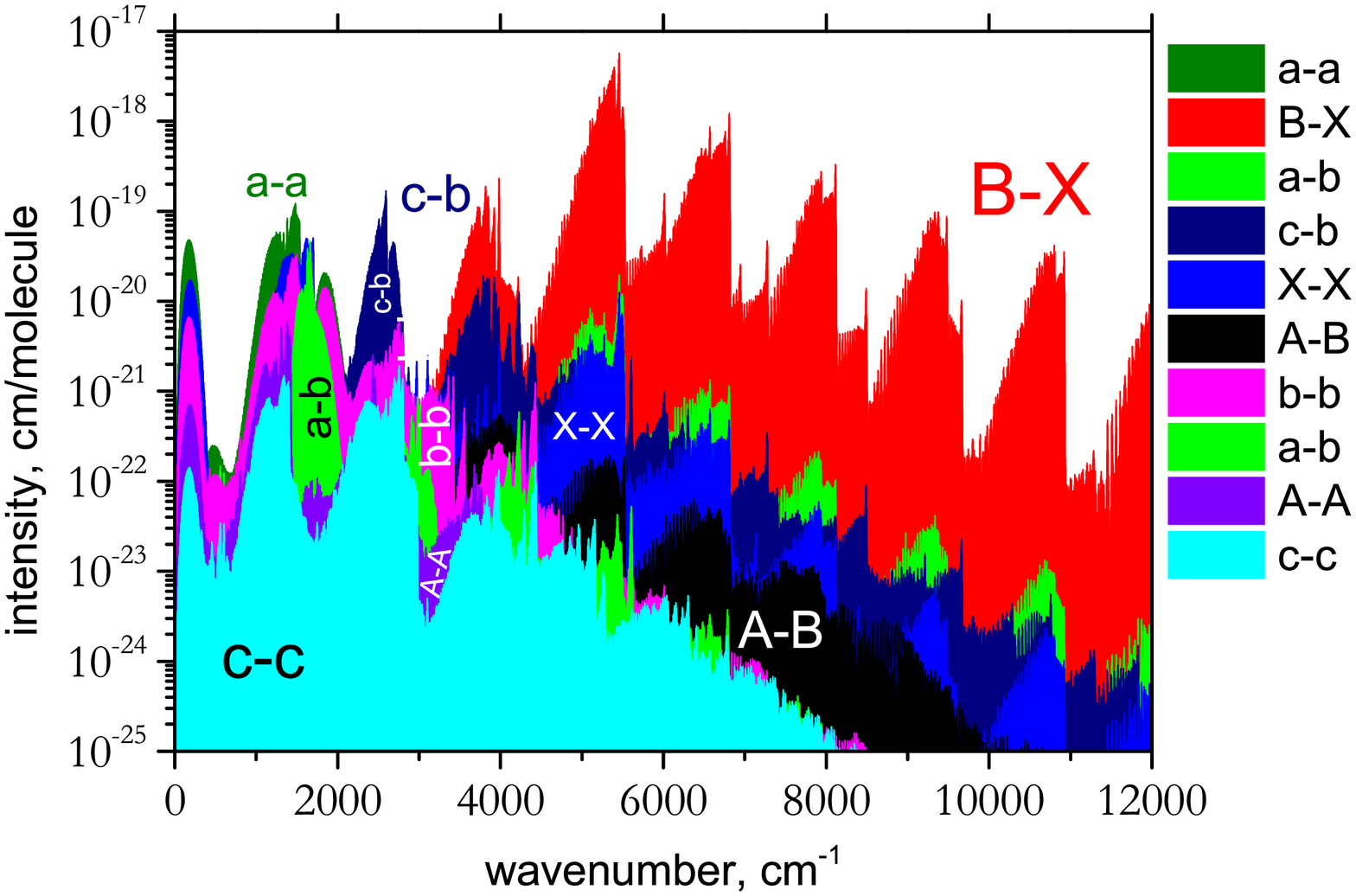}
\caption{Overview of the absorption line intensities of ScH at $T$=1500~K. \label{f:overview}}
\end{center}
\end{figure}

As an example, in fig.~\ref{f:298K} we show  absorption ($T=$ 298~K)
intensities of ScH generated using a stick-diagram. Figure~\ref{f:overview}
illustrates in detail the band structure of the absorption spectrum of ScH at
$T=1\,500$~K generated using a Gaussian line profile with the
half-width-at-half-maximum  (HWHM) of 5~\cm. As one can see, the strongest
features belong to the $X$--$X$, $B$--$X$, $a$--$a$, and $c$--$b$ electronic
bands. The triplet bands electronic bands $a$--$a$ and $c$--$b$ should be
strong enough to be potentially observable in the lab. 
Due to the spin-orbit couplings between different components forbidden bands also appear to contribute, with A–X and c–X being the strongest. These bands are significantly weaker than the dipole-allowed ones, but could still represent an important source of ScH opacity.

\begin{table}
\footnotesize
\caption{Sample extract from the energy file for $^{45}$ScH. The whole file contains
8~451 entries.}
\label{t:Energy-file}
\begin{tabular}{rrrrrrlrrrr c rrr}
\hline\hline
     $i$ & \multicolumn{1}{c}{$\tilde{E}$} &  $g$    & $J$  &  \multicolumn{1}{c}{$+/-$} &  \multicolumn{1}{c}{$e/f$} & State & $v$    & $|\Lambda|$& $|\Sigma|$ & $|\Omega|$& \\
\hline
\texttt{    1}&\texttt{     0.000000}&\texttt{ 16}&\texttt{  0}&\texttt{+   }&\texttt{f   }&\texttt{X1Sigma+    }&\texttt{    0}&\texttt{    0}&\texttt{    0}&\texttt{    0}\\
\texttt{    2}&\texttt{  1547.095548}&\texttt{ 16}&\texttt{  0}&\texttt{+   }&\texttt{f   }&\texttt{X1Sigma+    }&\texttt{    1}&\texttt{    0}&\texttt{    0}&\texttt{    0}\\
\texttt{    3}&\texttt{  3019.322250}&\texttt{ 16}&\texttt{  0}&\texttt{+   }&\texttt{f   }&\texttt{X1Sigma+    }&\texttt{    2}&\texttt{    0}&\texttt{    0}&\texttt{    0}\\
\texttt{    4}&\texttt{  3352.480112}&\texttt{ 16}&\texttt{  0}&\texttt{+   }&\texttt{f   }&\texttt{b3Pi        }&\texttt{    0}&\texttt{    1}&\texttt{    1}&\texttt{    0}\\
\texttt{    5}&\texttt{  4430.504406}&\texttt{ 16}&\texttt{  0}&\texttt{+   }&\texttt{f   }&\texttt{X1Sigma+    }&\texttt{    3}&\texttt{    0}&\texttt{    0}&\texttt{    0}\\
\texttt{    6}&\texttt{  4707.209870}&\texttt{ 16}&\texttt{  0}&\texttt{+   }&\texttt{f   }&\texttt{b3Pi        }&\texttt{    1}&\texttt{    1}&\texttt{    1}&\texttt{    0}\\
\texttt{    7}&\texttt{  5789.270187}&\texttt{ 16}&\texttt{  0}&\texttt{+   }&\texttt{f   }&\texttt{X1Sigma+    }&\texttt{    4}&\texttt{    0}&\texttt{    0}&\texttt{    0}\\
\texttt{    8}&\texttt{  6015.690883}&\texttt{ 16}&\texttt{  0}&\texttt{+   }&\texttt{f   }&\texttt{b3Pi        }&\texttt{    2}&\texttt{    1}&\texttt{    1}&\texttt{    0}\\
\texttt{    9}&\texttt{  7100.259438}&\texttt{ 16}&\texttt{  0}&\texttt{+   }&\texttt{f   }&\texttt{X1Sigma+    }&\texttt{    5}&\texttt{    0}&\texttt{    0}&\texttt{    0}\\
\texttt{   10}&\texttt{  7277.938899}&\texttt{ 16}&\texttt{  0}&\texttt{+   }&\texttt{f   }&\texttt{b3Pi        }&\texttt{    3}&\texttt{    1}&\texttt{    1}&\texttt{    0}\\
\texttt{   11}&\texttt{  8363.776161}&\texttt{ 16}&\texttt{  0}&\texttt{+   }&\texttt{f   }&\texttt{X1Sigma+    }&\texttt{    6}&\texttt{    0}&\texttt{    0}&\texttt{    0}\\
\texttt{   12}&\texttt{  8494.401968}&\texttt{ 16}&\texttt{  0}&\texttt{+   }&\texttt{f   }&\texttt{b3Pi        }&\texttt{    4}&\texttt{    1}&\texttt{    1}&\texttt{    0}\\
\texttt{   13}&\texttt{  9574.352490}&\texttt{ 16}&\texttt{  0}&\texttt{+   }&\texttt{f   }&\texttt{X1Sigma+    }&\texttt{    7}&\texttt{    0}&\texttt{    0}&\texttt{    0}\\
\texttt{   14}&\texttt{  9667.692015}&\texttt{ 16}&\texttt{  0}&\texttt{+   }&\texttt{f   }&\texttt{b3Pi        }&\texttt{    5}&\texttt{    1}&\texttt{    1}&\texttt{    0}\\
\texttt{   15}&\texttt{ 10718.414759}&\texttt{ 16}&\texttt{  0}&\texttt{+   }&\texttt{f   }&\texttt{b3Pi        }&\texttt{    6}&\texttt{    1}&\texttt{    1}&\texttt{    0}\\
\texttt{   16}&\texttt{ 10804.884934}&\texttt{ 16}&\texttt{  0}&\texttt{+   }&\texttt{f   }&\texttt{X1Sigma+    }&\texttt{    8}&\texttt{    0}&\texttt{    0}&\texttt{    0}\\
\texttt{   17}&\texttt{ 11788.395483}&\texttt{ 16}&\texttt{  0}&\texttt{+   }&\texttt{f   }&\texttt{b3Pi        }&\texttt{    7}&\texttt{    1}&\texttt{    1}&\texttt{    0}\\
\texttt{   18}&\texttt{ 11902.873682}&\texttt{ 16}&\texttt{  0}&\texttt{+   }&\texttt{f   }&\texttt{X1Sigma+    }&\texttt{    9}&\texttt{    0}&\texttt{    0}&\texttt{    0}\\
\texttt{   19}&\texttt{ 12788.251034}&\texttt{ 16}&\texttt{  0}&\texttt{+   }&\texttt{f   }&\texttt{b3Pi        }&\texttt{    8}&\texttt{    1}&\texttt{    1}&\texttt{    0}\\
\texttt{   20}&\texttt{ 12941.194344}&\texttt{ 16}&\texttt{  0}&\texttt{+   }&\texttt{f   }&\texttt{X1Sigma+    }&\texttt{   10}&\texttt{    0}&\texttt{    0}&\texttt{    0}\\
\texttt{   21}&\texttt{ 13714.254989}&\texttt{ 16}&\texttt{  0}&\texttt{+   }&\texttt{f   }&\texttt{b3Pi        }&\texttt{    9}&\texttt{    1}&\texttt{    1}&\texttt{    0}\\
\texttt{   22}&\texttt{ 13898.140315}&\texttt{ 16}&\texttt{  0}&\texttt{+   }&\texttt{f   }&\texttt{X1Sigma+    }&\texttt{   11}&\texttt{    0}&\texttt{    0}&\texttt{    0}\\
\texttt{   23}&\texttt{ 14551.459500}&\texttt{ 16}&\texttt{  0}&\texttt{+   }&\texttt{f   }&\texttt{b3Pi        }&\texttt{   10}&\texttt{    1}&\texttt{    1}&\texttt{    0}\\
\texttt{   24}&\texttt{ 14746.814942}&\texttt{ 16}&\texttt{  0}&\texttt{+   }&\texttt{f   }&\texttt{X1Sigma+    }&\texttt{   12}&\texttt{    0}&\texttt{    0}&\texttt{    0}\\
\hline

\end{tabular}

\mbox{}\\
$i$:   State counting number.     \\
$\tilde{E}$: State energy in \cm. \\
$g$: State degeneracy.            \\
$+/-$:   Actual state parity. \\
$e/f$:   Rotationless parity. \\
$v$:   State vibrational quantum number. \\
$|\Lambda|$:   Absolute value of $\Lambda$ (projection of the electronic angular momenum). \\
$|\Sigma|$:   Absolute value of $\Sigma$ (projection of the electronic spin). \\
$|\Omega|$:   Absolute value of $\Omega=\Lambda+\Sigma$ (projection of the
total angular momentum).

\end{table}

\begin{table}
\footnotesize
\caption{Sample extract from the transition file for $^{45}$ScH. The whole file contains
1~152~826 entries.}
\label{t:transition-file}
\begin{tabular}{rrr}
\hline
$f$& $i$ & \multicolumn{1}{r}{$A_{\rm if}$} \\
\hline
\texttt{ 1351}   & \texttt{ 1231}   &      \texttt{3.3006E-07}\\
\texttt{ 3574}   & \texttt{ 3468}   &      \texttt{3.8843E-07}\\
\texttt{ 5782}   & \texttt{ 5693}   &      \texttt{5.0269E-06}\\
\texttt{ 4942}   & \texttt{ 5037}   &      \texttt{9.9272E-06}\\
\texttt{ 7688}   & \texttt{ 7734}   &      \texttt{4.9607E-03}\\
\texttt{ 2070}   & \texttt{ 1952}   &      \texttt{3.8782E-01}\\
\texttt{ 2919}   & \texttt{ 2580}   &      \texttt{1.0196E+00}\\
\texttt{ 2804}   & \texttt{ 2692}   &      \texttt{1.0196E+00}\\
\texttt{ 1362}   & \texttt{ 1713}   &      \texttt{8.5042E-08}\\
\texttt{ 5638}   & \texttt{ 5727}   &      \texttt{6.8009E-04}\\
\texttt{ 3137}   & \texttt{ 3242}   &      \texttt{4.2340E-06}\\
\texttt{ 3672}   & \texttt{ 3568}   &      \texttt{1.9332E-04}\\
\texttt{ 7406}   & \texttt{ 7350}   &      \texttt{4.6729E-06}\\
\texttt{ 5984}   & \texttt{ 5901}   &      \texttt{1.2467E-05}\\
\texttt{ 3396}   & \texttt{ 3505}   &      \texttt{9.4844E-05}\\
\texttt{  993}   & \texttt{ 1110}   &      \texttt{1.0904E-06}\\
\texttt{ 3324}   & \texttt{ 2999}   &      \texttt{1.9023E-05}\\
\texttt{ 2398}   & \texttt{ 2282}   &      \texttt{1.0987E-04}\\
\hline
\end{tabular}

\mbox{}\\
$f$: Upper (final) state counting number. \\
$i$: Lower (initial) state counting number.\\
$A_{\rm if}$:  Einstein~A coefficient in s$^{-1}$.

\end{table}


\section{Conclusions}
A hot line list containing pure rotational, ro-vibrational and vibronic
transitions for ScH was generated using new {\it ab initio} potential energy,
dipole moment, spin-orbit, and electronic angular moment curves obtained at a
high level of theory. The work was performed with a view to astrophysical
applications. The analysis of the importance of different absorption bands for
the opacities of ScH is presented.

The complexity of the electronic structure problem when transition metals are
involved means that the accuracy of these calculations is much worse than what
is normally achievable for small molecules containing light main-group
elements. To help mitigate this problem, it is desirable to utilise
experimental data to improve the accuracy of the results. We have done this for
the singlet states.

However,  for transitions involving triplet states there are no measured data
for us to use: for example, the recent triplet-triplet measurements by Mukund
{\it et al} \cite{14MoBhNa.ScH} in the 17~940 \cm\ region lie well above our
calculated transitions and have not reported any absolute energies of the lower
$a^{}\leftexp{3}\Delta$ electronic state. Besides, their only 
parameter with which we can directly compare, $B_e$, appears to be too low to be fully trusted.  Thus for the
triplet states we used the $T_e$  valued taken from our {\it ab initio}
calculations, which may be in error by up to a few thousands \cm, implying that
the entire bands may be erroneously shifted by similar amounts. Also, structure
due to perturbations in both the singlet and triplet manifolds are dependent
on the potential energy curve separations and therefore may not be reproduced
accurately. However we expect that our line list to be quantitatively accurate
enough for the singlet state energies and to provide a detailed spectral
structure of individual bands to be useful.

\section*{Acknowledgements}

This work was supported by ERC Advanced Investigator Project 267219.

\bibliographystyle{tMPH}

\end{document}